\documentclass[aps,prl,showpacs,twocolumn,superscriptaddress]{revtex4}
\usepackage{amsmath}
\usepackage{amssymb}
\usepackage{epsfig}
\usepackage{color}
\usepackage{amsmath}
\usepackage[colorlinks]{hyperref}
\usepackage{graphicx,amsmath}
\usepackage[normalem]{ulem}
\usepackage{indentfirst}
\UseRawInputEncoding \input
\hypersetup{colorlinks,citecolor=red,linkcolor=blue,urlcolor=blue}
\usepackage{graphicx,amsmath}

\begin{document}
\title{Ultra spin liquid in $\rm Lu_3Cu_2Sb_3O_{14}$}
\author{V. R. Shaginyan}\email{vrshag@thd.pnpi.spb.ru}
\affiliation{Petersburg Nuclear Physics Institute, NRC Kurchatov
Institute, Gatchina, 188300, Russia}\affiliation{Clark Atlanta
University, Atlanta, GA 30314, USA} \author{A. Z. Msezane}
\affiliation{Clark Atlanta University, Atlanta, GA 30314, USA}
\author{S. A. Artamonov}\affiliation{Petersburg Nuclear Physics
Institute, NRC Kurchatov Institute, Gatchina, 188300,
Russia}\author{G. S. Japaridze} \affiliation{Clark Atlanta
University, Atlanta, GA 30314, USA}
\author{Y. S. Leevik} \affiliation{National Research University
Higher School of Economics, St.Petersburg, 194100, Russia}

\begin{abstract}
We analyze recent measurements of $C/T$,  specific heat $C$ divided
by temperature $T$, of recently observed ultra spin liquid. The
measurements are carried out in magnetic fields on the triangular
lattice compound $\rm Lu_3Cu_2Sb_3O_{14}$. We show that the
obtained heat capacity $C_{\rm mag}$ formed by ultra spin liquid as
a function of temperature $T$ versus magnetic field $B$ behaves
very similar to the electronic specific heat $C_{el}$ of the HF
metal $\rm YbRh_2Si_2$ and the quantum magnet $\rm
ZnCu_3(OH)_6Cl_2$. We demonstrate that the effective mass of spinon
$M^*\propto C_{\rm mag}/T$ exhibits the universal scaling
coinciding with scaling observed in heavy fermion (HF) metals and
in $\rm ZnCu_3(OH)_6Cl_2$. Based on these observations we conclude
that a strongly correlated spin liquid determines the thermodynamic
properties of the ultra spin liquid of $\rm Lu_3Cu_2Sb_3O_{14}$.
\end{abstract}


\maketitle

\section{Introduction}

By now, a number of quantum spin liquids (QSL) with various types
of ground states are proposed
\cite{and,herb3,herb,balents:2010,sci_slg,mend2011,han:2012,shagqsl,varma}.
It is expected that QSLs define the thermodynamic, transport and
relaxation properties of frustrated magnets and represent the new
state of matter formed by strongly correlated Fermi systems
\cite{shagqsl,book20}. Some of QSLs are formed with fermionic
quasiparticles with the effective mass $M^*$ which we call spinons.
Spinons carry spin $\sigma=1/2$ and no charge. At temperature $T=0$
the Fermi sphere is comprised from spinons with the Fermi momentum
$p_F$. The Fermi sphere can be located near the topological Fermion
condensation phase transition (FCQPT) that forms flat bands
\cite{ks,vol,prl,book,book20,Khod2020}. As a result, the strongly
correlated quantum spin liquid (SCQSL) emerges that allows one to
describe numerous data related to the thermodynamic, relaxation and
transport properties of frustrated magnetic insulators
\cite{shagqsl,prl,book,book20,shaginyan:2011,shaginyan:2011:C,shaginyan:2012:A}.
Recent measurements of the specific heat of the triangular lattice
compound $\rm Lu_3Cu_2Sb_3O_{14}$ performed in the pretense of
magnetic field yield  important experimental facts shedding light
on the nature of newly discovered ultra spin liquid \cite{varma}.
It is suggested that the ultra spin liquid is formed by quantum
fluctuations. In turn these fluctuations lead to either a large
specific heat peak from singlet excitations, or to a degenerate
topological singlet ground state \cite{varma}.

In this Letter we analyze measurements in magnetic fields of the
specific heat $C_{\rm mag}/T$ of the ultra spin liquid \cite{varma}
that forms the thermodynamic properties of $\rm
Lu_3Cu_2Sb_3O_{14}$. We show that obtained $C_{\rm mag}/T\propto
M^*$ as a function of two independent variables magnetic field $B$
and temperature $T$ behaves very similar to the electronic specific
heat $C_{el}/T$ of the heavy fermion (HF) metal $\rm YbRh_2Si_2$
and the specific heat $C_{\rm mag}/T$ of the quantum magnet $\rm
ZnCu_3(OH)_6Cl_2$. This similarity allows us to conclude that the
corresponding QSL is represented by spinons that fill in the Fermi
sphere with the Fermi momentum $p_F$, while the thermodynamic
properties of the ultra spin liquid is determined by the strongly
correlated quantum spin liquid, rather than by fluctuations. We
demonstrate that the effective mass of spinon $M^*\propto C_{\rm
mag}/T$ exhibits scaling similar to that of $C_{el}/T$ and $C_{\rm
mag}/T$.

\section{Universal scaling behavior of the ultra spin liquid}

The spinons of the triangular lattice compounds occupy a symmetric
positions. Consequently,the ground state energy weakly depends on
the spins configuration. As a result, the triangular lattice is
close to a topologically protected flat branch of the spectrum with
zero excitation energy
\cite{green:2010,heikkila:2011,shaginyan:2011,pr,book}. Therefore,
the topological FCQPT can be considered as a quantum critical point
(QCP) of the $\rm Lu_3Cu_2Sb_3O_{14}$ ultra spin liquid. We assume
that the elementary magnetic excitations, dubbed spinons, defining
the thermodynamic properties, carry the effective mass $M^*$, zero
charge and spin $\sigma=1/2$ and occupy the corresponding Fermi
sphere with the Fermi momentum $p_F$. They form the excitation
spectrum typical for HF liquid located near the topological FCQPT
and represent HF quasiparticles of deconfined QSL. The ground state
energy $E(n)$ is given by the Landau functional that depends on the
spinon distribution function $n_\sigma({\bf p})$, with ${\bf p}$
being the momentum. Near the FCQPT point, the effective mass $M^*$
is governed by the Landau equation \cite{land,pr}
\begin{eqnarray}
\label{HC3} &&\frac{1}{M^*(T,B)}=\frac{1}{M^*(T=0,B=0)}\\&+&
\frac{1}{p_F^2}\sum_{\sigma_1}\int\frac{{\bf p}_F{\bf p_1}}{p_F}
F_{\sigma,\sigma_1}({\bf p_F},{\bf p}_1)\frac{\partial\delta
n_{\sigma_1}({\bf p}_1)} {\partial{p}_1}\frac{d{\bf p}_1}{(2\pi)
^3}.\nonumber
\end{eqnarray}
Note that both functional $E(n)$ and Eq. \eqref{HC3} are exact
\cite{book20,pla}.  This fact makes firm ground to construct the
theory of HF compounds \cite{book,pr,book20}. Considering that, due
to the geometric frustration, QSL in $\rm Lu_3Cu_2Sb_3O_{14}$
\cite{varma} is located near the topological FCQPT, we use the
theory of HF compounds to describe SCQSL of $\rm
Lu_3Cu_2Sb_3O_{14}$. This theory allows quantitative analysis of
the thermodynamic, relaxation and transport properties of both HF
compounds containing QSL and HF metals
\cite{shaginyan:2011,book,pr,book20,shagqsl,prl}. As we shall see,
the thermodynamic properties of QSL in question coincide with those
of SCQSL in the frustrated magnet $\rm ZnCu_3(OH)_6Cl_2$ and HF
electrons in the HF metal $\rm YbRh_2Si_2$.

\begin{figure} [! ht]
\begin{center}
\includegraphics [width=0.55\textwidth]{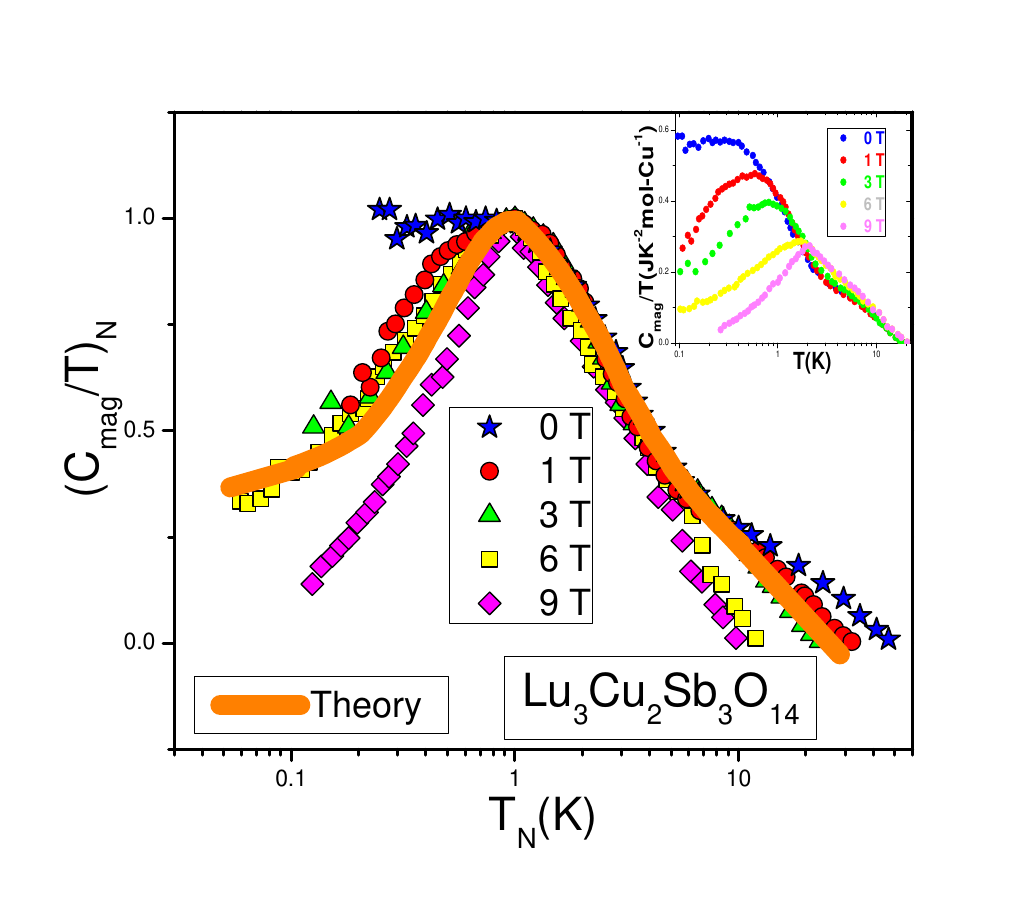}
\end{center}
\caption{(color online). The normalized specific heat $(C_{\rm
mag}/T)_N$ of $\rm Lu_3Cu_2Sb_3O_{14}$ as a function of the
normalized temperature $T_N$ measured in the presence of the
magnetic field. The normalized specific heat $C_{\rm mag}/T$ is
extracted from the measurement of the specific heat of $\rm
Lu_3Cu_2Sb_3O_{14}$ \cite{varma} shown in the inset. The solid
orange curve corresponds to our theoretical calculations based on
Eq. \eqref{HC3}. The same curve is depicted in Figs. \ref{fig1} and
\ref{fig4}, demonstrating the universal scaling behavior of the
thermodynamic properties of the ultra spin liquid.}\label{fig1p}
\end{figure}

In Eq. \eqref{HC3} we rewrite the spinon distribution function as
$\delta n_{\sigma}({\bf p})\equiv n_{\sigma}({\bf
p},T,B)-n_{\sigma}({\bf p},T=0,B=0)$. The Landau interaction drives
the system to the FCQPT point, where the Fermi surface alters
topology of the system so that the effective mass acquires strong
temperature and the field dependence \cite{pr,ckz,khodb,book20}, as
seen from the inset of Fig. \ref{fig1p}. Indeed, it is seen from
the inset that the effective mass $M^*(T,B)\propto C_{\rm mag}/T$
strongly depends on both $T$ and $B$.
At the topological FCQPT the term $1/M^*(T=0,B=0)$ vanishes and Eq.
\eqref{HC3} becomes homogeneous and therefore is solved
analytically. At $B=0$, the effective mass depends on $T$
exhibiting the non-Fermi liquid (NFL) behavior \cite{pr}
\begin{equation}
M^*(T)\simeq a_TT^{-d}.\label{MTT}
\end{equation}
Here $d=2/3$ or $d=1/2$, and can be defined from the corresponding
experimental facts \cite{shagqsl,book20}. In the case of $\rm
Lu_3Cu_2Sb_3O_{14}$ we have $d=2/3$. At finite $T$ magnetic field
$B$ drives the system to the Landau Fermi liquid (LFL) behavior
with
\begin{equation}
M^*(B)\simeq a_BB^{-d}.\label{MBB}
\end{equation}
Here $a_T$ and $a_B$ are fitting parameters.

At finite $B$ and $T$ near the topological FCQPT, the solution of
Eq. \eqref{HC3} $M^*(B,T)$ is approximated by a universal
interpolating function \cite{pr}. The interpolation links the LFL
($M^*(T)\propto const)$ and NFL ($M^*(T)\propto T^{-2/3}$) regions.
\begin{figure} [! ht]
\begin{center}
\includegraphics [width=0.47\textwidth]{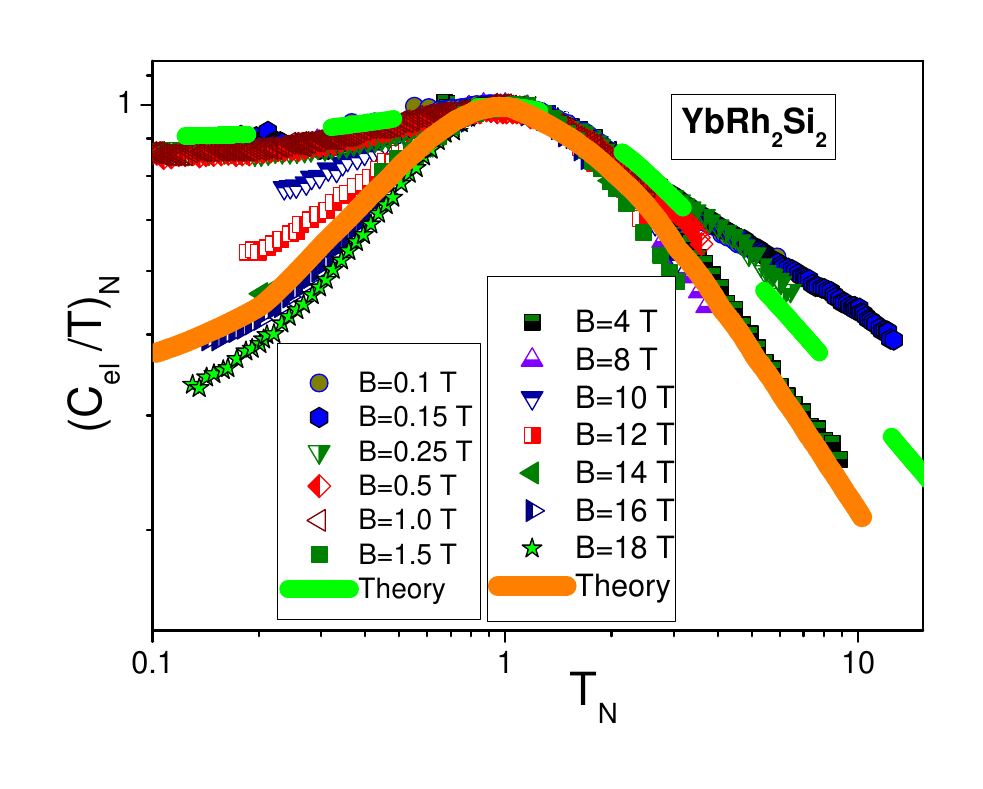}
\end{center}
\caption{(color online). The normalized specific heat $(C_{\rm
el}/T)_N=M^*_N$ of $\rm YbRh_2Si_2$ versus normalized temperature
$T_N$ as a function of $B$ at low (the left hand box symbols) and
high (the right hand box symbols) magnetic fields, extracted from
the measurement of the specific heat on the archetypical HF metal
$\rm YbRh_2Si_2$ \cite{gegenwart:2006:A,oeschler:2008}. The
low-field calculations are depicted by the solid dashed line
tracing the scaling behavior of $M^*_N$. The high-field
calculations (solid orange line) are performed for high magnetic
fields $B\sim 18$ T when the quasiparticle band becomes fully
polarized \cite{shaginyan:2011:C}. At magnetic fields $B\geq 4$ T
the specific heat exhibits the same behavior as $(C_{\rm
mag}/T)=M^*_N$ of $\rm Lu_3Cu_2Sb_3O_{14}$ shown in Fig.
\ref{fig1p}.}\label{fig1}
\end{figure}

{It is seen from the inset of Fig. \ref{fig1p} that $C_{\rm mag}/T$
reaches its maximum value $(C_{\rm mag}/T)_{\rm max}(B)$ under the
application of magnetic fields at some temperature $T_{\rm
max}(B)$. To reveal the scaling behavior, we introduce the
dimensionless normalized specific heat $(C_{\rm mag}/T)_N$ as a
function of the dimensionless normalized temperature $T_N=T/T_{\rm
max}(B)$ \cite{pr}
\begin{equation}
(C_{\rm mag}/T)_N=\frac{C_{\rm mag}/T}{(C_{\rm mag}/T)_{\rm
max}}.\label{norm}
\end{equation}
As seen from Fig. \ref{fig1p}, that $(C_{\rm mag}/T)_N(T_N)$ as a
function of $T_N$ merges into a single curve independent of the
applied magnetic field.}

{In the same way, to construct the interpolating equation and
reveal the universal scaling behavior of the effective mass
$M^*\propto C_{\rm mag}/T$, we use the dimensionless normalized
effective mass $M^*_N$ and the dimensionless normalized temperature
$T_N$, defined by dividing the effective mass $M^*(T,B)$ by its
maximal values, $M^*_{\rm max}(T,B)$, and temperature $T$ by
$T_{\rm max}$ at which the maximum $M^*_N$ occurs, $T_N=T/T_{\rm
max}$ \cite{pr}. Magnetic field $B$ appears in Eq. \eqref{HC3} only
in the combination $\mu_BB/k_BT$, thus $k_BT_{\rm max}\simeq
\mu_BB$ where $k_B$ is the Boltzmann constant and $\mu_b$ is the
Bohr magneton \cite{ckz,pr}. Thus, $T_{\rm max}\propto B$ and
$T_N\propto T/B$, see e.g. \cite{pr,shag20}. The normalized
effective mass $M^*_N=M^*/M^*_{\rm max}(T_N\propto T/B)=(C_{\rm
mag}/T)_N$ is given by the interpolating function that approximates
the solution of Eq. \eqref{HC3}\, \cite{pr}}
\begin{equation}M^*_N(y)\approx c_0\frac{1+c_1y^2}{1+c_2y^{8/3}}.
\label{UN2}
\end{equation}
Here $c_0=(1+c_2)/(1+c_1)$, $c_1$ and $c_2$ are fitting parameters,
and $y=T/T_{\rm max}\propto T/B$. It is seen from Eq. \eqref{UN2}
that under the application of magnetic field $M^*$ becomes finite
and at low temperatures the system exhibits the LFL behavior
$C_{\rm mag}(T)/T\propto M^*(T)\simeq M^*(T=0)+a_1T^2$. As seen
from the inset of Fig. \ref{fig1p}, at increasing temperatures
$M^*\propto C_{\rm mag}/T$ increases  and enters the crossover
region, reaching its maximum $M^*_{\rm max}\propto (C_{\rm
mag}(T)/T)_{\rm max}$ at $T=T_{\rm max}$, with subsequent
diminishing given by Eqs. \eqref{MTT} and \eqref{UN2}. {Equation
\eqref{UN2} exhibits the scaling behavior, demonstrating the vivid
property of Eq. \eqref{HC3} at the topological FCQPT: the function
$M^*(T,B)$ of two variables becomes the function $M^*$ of the
single variable $T_N\propto T/B$.} We utilize Eq. \eqref{UN2} to
outline the universal scaling behavior and clarify our calculations
based on Eq. \eqref{HC3}.
\begin{figure} [! ht]
\begin{center}
\includegraphics [width=0.47\textwidth]{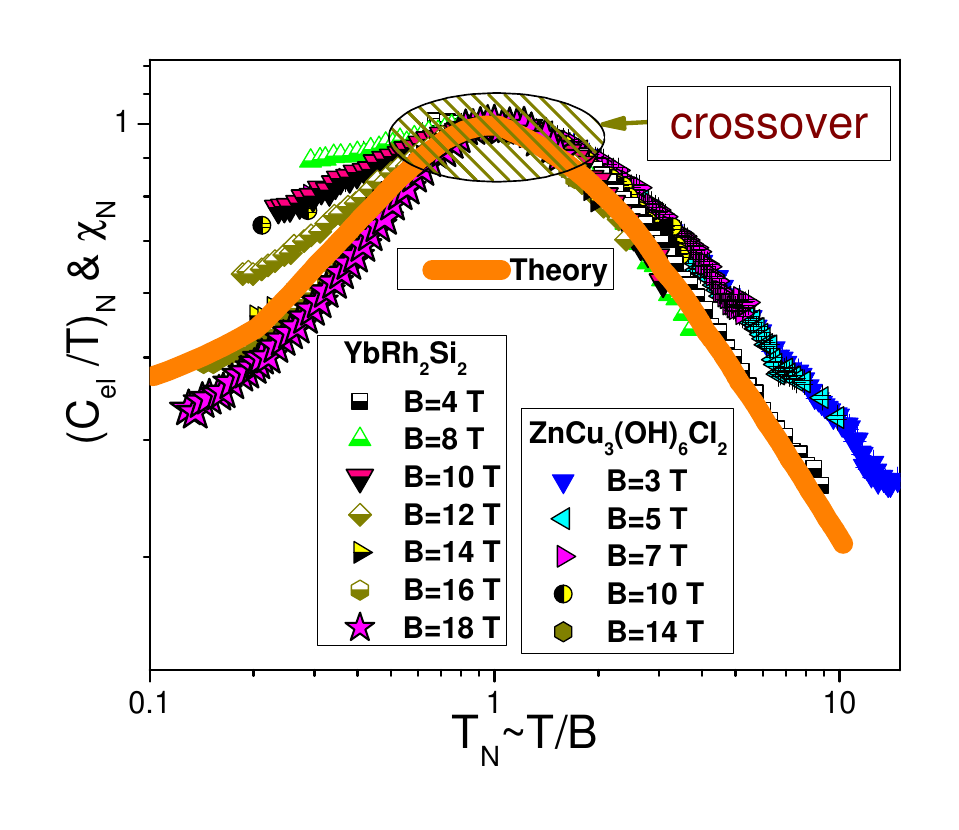}
\end{center}
\caption{(color online). Normalized magnetic susceptibility
$\chi_N=\chi/\chi_{\rm max}=M^*_N$ versus normalized temperature
$T_N\propto T/B$ as a function of magnetic field shown in the
legend. The data are extracted from the measurements of the
magnetic susceptibility $\chi(T,B)$ on $\rm ZnCu_3(OH)_6Cl_2$
\cite{herb3}. Normalized data of $(C_{\rm el}/T)_N=M^*_N$ are
obtained from the specific heat $C_{\rm el}/T$ of $\rm YbRh_2Si_2$
measured in the presence of magnetic field $B$ (the legend)
\cite{gegenwart:2006:A}. The solid curve corresponds to the
theoretical calculations at $B\simeq 18$ T when the quasiparticle
band is fully polarized. The curve represents the universal scaling
behavior of $M^*_N$ coinciding with the scaling of the ultra spin
liquid which is depicted in Fig. \ref{fig1p}. The crossover from
the LFL behavior to the NFL one is shown by the arrow.}
\label{fig4}
\end{figure}

The scaling of $(C_{\rm mag}/T)_N=M_N$, extracted from $C_{\rm
mag}(T,B)/T$ \cite{varma}, is reported in Fig. \ref{fig1p}. The
data for a wide range of $B$ up to 9 T merge well into a single
curve. Figure \ref{fig1} reports the normalized specific heat
$(C_{\rm el}/T)_N=M^*_N$ of $\rm YbRh_2Si_2$ versus normalized
temperature $T_N$ as a function of $B$. It is seen that at low
$T_N\lesssim0.1$ normalized specific heat $(C_{\rm el}/T)_N\simeq
0.4$ \cite{shaginyan:2011:C}. This value is determined by the
polarization of the heavy electron liquid in magnetic fields $B>4$
T, and coincides with that of $\chi_N=M^*_N$ obtained on $\rm
ZnCu_3(OH)_6Cl_2$ and shown in Fig. \ref{fig4}. Results of our
calculations are presented by the same solid orange curve in Figs.
\ref{fig1p}, \ref{fig1} and \ref{fig4}. Note that at low $T_N$ and
at low magnetic field, when the polarization is negligible,
$(C_{\rm el}/T)_N\simeq 0.9$, as seen from Fig. \ref{fig1}. Thus,
the behavior of $C_{\rm mag}/T$ reported in Fig. \ref{fig1p} is of
universal character, for we observe that $(C_{\rm
mag}/T)_N=M^*_{N}$ of $\rm Lu_3Cu_2Sb_3O_{14}$ behaves like
$(C_{\rm el}/T)_N=\chi_N=M^*_N$ shown in Figs. \ref{fig1} and
\ref{fig4}. The data shown in these Figs are extracted from
measurements on $\rm ZnCu_3(OH)_6Cl_2$ and $\rm YbRh_2Si_2$
\cite{herb3,gegenwart:2006:A,oeschler:2008} As a result, the ultra
spin liquid can be viewed as SCQSL that exhibits gapless behavior
in a magnetic field. We note that the data at magnetic field $B=0$
and $T_N<1$ indicate that the ultra spin liquid exhibits the LFL
behavior. Otherwise, the spin liquid, being on the ordered side of
FCQPT, would have been consumed by phase transition, eliminating
the corresponding entropy excess \cite{pr,book20}. Therefore, we
expect that a stable QSL should be close to the topological FCQPT
located on the disordered side of the phase transition.
\begin{figure} [! ht]
\begin{center}
\includegraphics [width=0.50\textwidth]{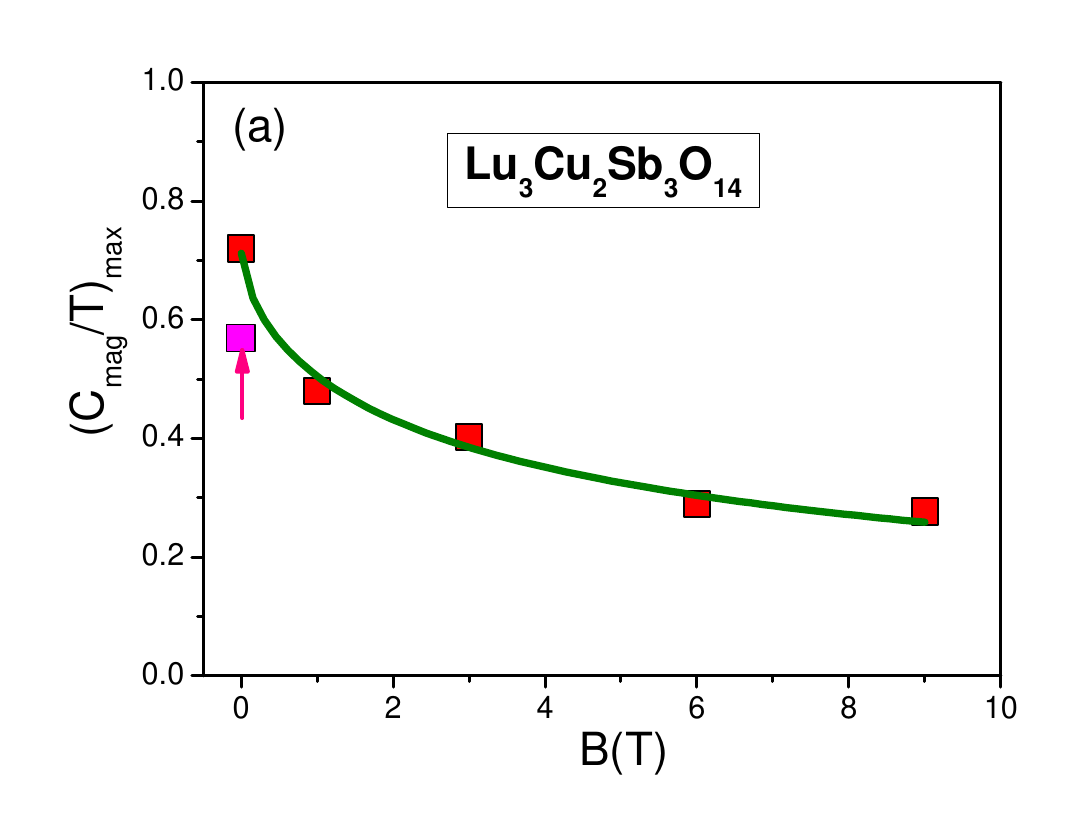}
\includegraphics [width=0.50\textwidth]{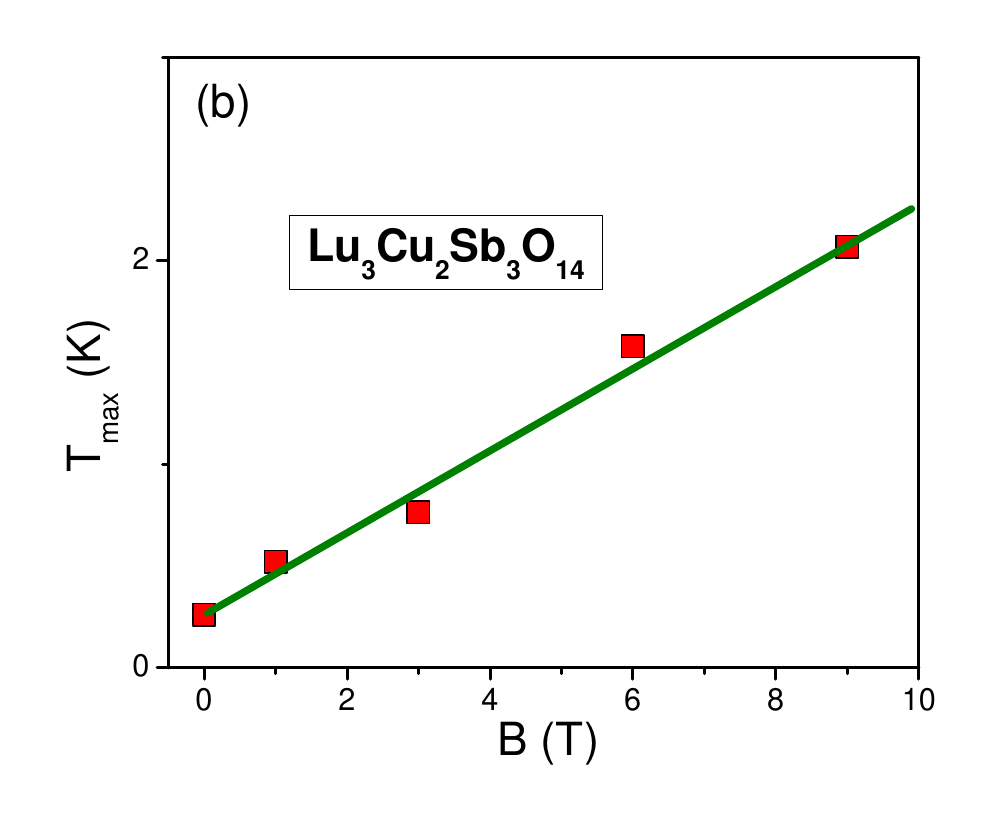}
\end{center}
\caption{(color online). (a) The maximums $(C_{\rm mag}/T)_{\rm
max}$ of $C_{\rm mag}/T$ versus magnetic field $B$ are shown by the
solid squares, see the inset, Fig. \ref{fig1p}. The solid curve is
approximated by $M^*_{\rm max}(B)\propto B^{-2/3}$, see Eq.
(\ref{MBB}). The arrow shows the position of $(C_{\rm mag}/T)_{\rm
max}$ at $B=0$ with the subtracted impurity Schottky contribution
\cite{varma}. (b) The temperature $T_{\rm max}(B)$, at which the
maximums of $(C_{\rm mag}/T)$ are located, see Fig. \ref{fig1} (a).
The solid straight line represents the function $T_{\rm max}\propto
B$.}\label{fig5}
\end{figure}

In Fig. \ref{fig5} (a), the solid squares denote the values of the
maxima $(C_{\rm mag}/T)_{\rm max}(B)$, taken from the inset of Fig.
\ref{fig1p}. In Fig. \ref{fig5} (a), the corresponding values of
$T_{\rm max}(B)$ are shown versus magnetic field $B$. It is seen
that the agreement between the theory (solid curve) and the
experiment is good. At $B=0$ the arrow shows the position of the
maximum with the subtracted impurity Schottky contribution
\cite{varma}. We believe that there is no reason to subtract the
the impurity Schottky contribution, as it is done in Ref.
\cite{varma}, since both the impurities and the pure crystal
holding SCQSL form the integral SCQSL \cite{shagqsl}. In Fig.
\ref{fig5} (b), the solid straight line displays the function
$T_{\rm max}(B)$. It is seen that at all temperatures the data are
well approximated by the straight line. Note that at $B=0$ the
$T_{\rm max}$ is not clearly discriminated, nonetheless at $B=0$
and $T\to0$ the specific heat $C_{\rm mag}/T$ demonstrates the LFL
behavior, as seen from the inset of Fig. \ref{fig1p}. This behavior
indicates that SCQSL of $\rm Lu_3Cu_2Sb_3O_{14}$ is not exactly
placed at FCQPT and at $T\to 0$ the system exhibits the LFL
behavior and the absence of a gap. This conclusion concurs with the
general properties of the phase diagrams of HF metals and quantum
magnets \cite{shagqsl,shag2014}. To clarify the mentioned above
properties, one needs to carry out low temperature measurements of
both the magnetic susceptibility $\chi$ and the thermal transport
under the application of magnetic fields. A few remarks are in
order here. Recent measurements of the low-temperature thermal
conductivity $\kappa$ have shown that the value of $\kappa(T\to0)$
strongly depends on the disorder of quantum magnet and at high
disorder $\kappa(T\to0)\to 0$,  see e.g. \cite{yamash20,prr}. Such
a behavior with $\kappa(T\to0)\to 0$ can signal that QSL is not
presented. On the other hand, the thermodynamic properties of some
quantum magnets with $\kappa(T\to0)\to 0$ demonstrate the typical
behavior of HF metals and one is to suggest that the thermodynamic
properties of these magnets are defined by QSL
\cite{shaginyan:2011,shagqsl,book20}. Thus, we have to assume that
there are presented at least the two type of QSL: one of them is
presented by QSL with high resistance to the heat transport, that
is $\kappa(T\to0)\to 0$, and another is characterized by
$\kappa(T\to0)$ being finite. In the latter case $\kappa$ depends
on magnetic field resembling the magnetoresistance of HF metals
\cite{shaginyan:2011:C,shaginyan:2012:A}. We speculate that in two
dimensional systems, formed by the kagome lattice, spinons can
participate in weakly bound states with impurities and that bound
states strongly obstruct the heat transport.

Perspective materials, where QSL can be presented, are the Kitaev
materials. They can be broadly defined as Mott insulators that
exhibit specific exchange interactions and are thought to have
unconventional forms of magnetism like QSLs \cite{rvb2}. The
experimentally studied examples are $\rm Na_2IrO_3$, $\rm
\alpha-Li_2IrO_3$ and $\rm\alpha-RuCl_3$ where local moments are
aligned in interacting hexagonal layers, see e.g.
\cite{savary,shagqsl,nat18,nat21}. Measurements of thermal
conductivity $\kappa(B)$ under the application of magnetic field
$B$ on the insulator $\rm\alpha-RuCl_3$ have shown that $\kappa(B)$
is finite at $T\to0$ when the antiferromagnetic order is suppressed
by magnetic field $B=B_c\simeq 7$ T, while $\kappa(B)$ is a growing
function at $B>B_c$ \cite{nat18,nat21}, since $\kappa(B)\propto
(M^*(B))^{-2}$ with $M^*(B)$ is given by Eq. \eqref{MBB}
\cite{shag13}. These observations are in good agreement with the
behavior of $\kappa(B)$ of SCQSL, and allows us to suggest that QSL
of $\rm\alpha-RuCl_3$ represents SCQSL, resembling the
corresponding behavior of HF metals
\cite{shagqsl,book20,shaginyan:2012:A,shag13}. A detailed
consideration of these items will be published elsewhere.

\section{Schematic phase diagram of $\rm Lu_3Cu_2Sb_3O_{14}$}

\begin{figure}[!ht]
\begin{center}
\includegraphics [width=0.47\textwidth]{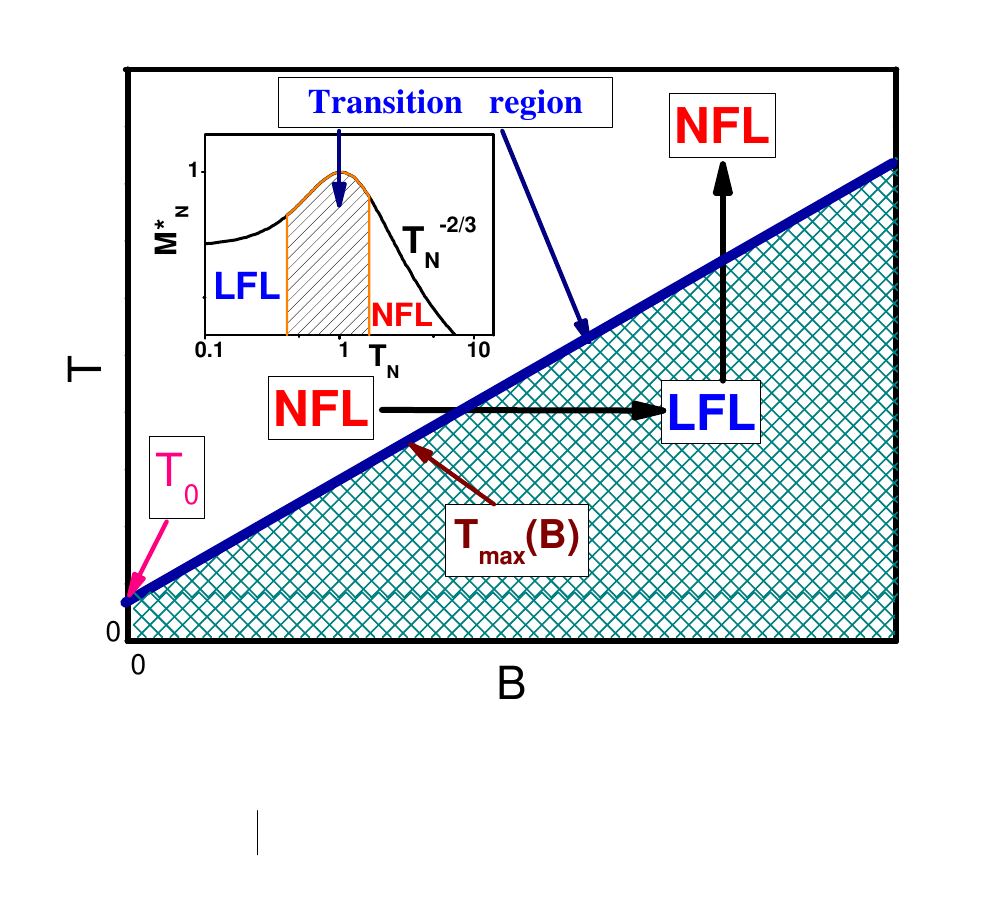}
\end{center}
\vspace*{-2.0cm} \caption{(color online). Schematic $T-B$ phase
diagram of $\rm Lu_3Cu_2Sb_3O_{14}$ with magnetic field as the
control parameter. The vertical and horizontal arrows depicts
LFL-NFL transitions at fixed $B$ and $T$, respectively. The line
shown by the arrow represents the crossover region at $T_{\rm
max}(B)$, and displays the function $T_{\rm max}(B)$. $T_0$ is the
temperature at which the LFL behavior occurs. The inset shows a
plot of the normalized effective mass $M^*_N$ versus the normalized
temperature $T_N$. Transition region, at which $M^*$ reaches its
maximum $M^*_{\rm max}$ at $T_N=T/T_{\rm max}=1$, is shown by the
arrow, see the inset of Fig. \ref{fig1p} as well.}\label{fg6}
\end{figure}

The schematic phase diagram of $\rm Lu_3Cu_2Sb_3O_{14}$, based on
Eq. \eqref{UN2} and Fig. \ref{fig1p}, is depicted in Fig.
\ref{fg6}. At $T=0$ and $B=0$ the system is located before the
topological FCQPT, that is on its disordered side. Therefore, at
$T=T_0$ the system exhibits the LFL behavior. Both magnetic field
$B$ and temperature $T$ play the role of the control parameters,
shifting the system from its location close  to the topological
FCQPT and driving it from the NFL to LFL regions as shown by the
vertical and horizontal arrows. At fixed temperatures the increase
of $B$ drives the system from the NFL to the LFL region.
This behavior is seen from Fig. \ref{fig5} (b): Under the
application of $B$, $T_{\rm max}$ shifts to higher temperatures.
Indeed, at $T<T_{\rm max}$ the system exhibits the LFL behavior
\cite{pr}. On the contrary, at fixed $B$ and growing temperatures
$T$, the system goes along the vertical arrow from the LFL to NFL
region. The inset to Fig. \ref{fg6} displays the behavior of the
normalized effective mass $M^*_N$ versus the normalized temperature
$T_N\propto T/B$ that follows from Eq. \eqref{UN2}. It is seen that
the region $T_N\sim 1$ represents the crossover region between the
LFL behavior with almost constant effective mass and the NFL
behavior, exhibiting the $T^{-2/3}$ dependence, see Eq. \eqref{MTT}
and Fig. \ref{fig4}. {It is worthy noting that in the framework of
the theory of Fermion condensation it is possible to explain the
crossover from the NFL behavior to LFL one under the application of
tiny magnetic field \cite{pr,shagqsl,book,book20,shag16}, while the
application of pressure does not change the NFL behavior, see e.g.
\cite{s2015}. Such a behavior observed in the heavy-fermion
superconductor $\rm \beta-YbAlB_4$, representing a strange metal
located away from a magnetic instability, is not accompanied by
fluctuations \cite{s2015}. In the same way, one cannot employ
quantum fluctuations to explain the corresponding properties of the
phase diagram \ref{fg6} and the dependencies displayed in Fig.
\ref{fig5} (a,b) \cite{varma}.} Thus, the general features of the
schematic phase diagram \ref{fg6} demonstrate that the
thermodynamic properties of $\rm Lu_3Cu_2Sb_3O_{14}$ are close to
those of the HF metal $\rm YbRh_2Si_2$ and $\rm ZnCu_3(OH)_6Cl_2$
\cite{pr,shagqsl,book20}. This confirms our observation that the
ultra spin liquid of $\rm Lu_3Cu_2Sb_3O_{14}$ is represented by
SCQSL.

\section{Summary}

In summary, we have shown that the ultra spin liquid of $\rm
Lu_3Cu_2Sb_3O_{14}$ can be viewed as a strongly correlated Fermi
system whose  thermodynamic properties are defined by SCQSL located
near FCQPT. Our calculations of the specific heat $C_{\rm mag}/T$
and the constructed phase diagram are in a good agreement with the
experimental data. The revealed scaling of $C_{\rm mag}/T$
coincides with that observed in the HF metal $\rm YbRh_2Si_2$ and
quantum magnet $\rm ZnCu_3(OH)_6Cl_2$. We have also explained the
strong dependence of the maximums of $C_{\rm mag}/T$ on magnetic
field $B$. We remark that such a behavior can hardly be explained
within theories based on different kinds of fluctuations. Thus, the
ultra spin liquid is well represented by SCQSL, and, therefore, can
be well described within the theory of Fermion condensation
\cite{pr,shagqsl,book,book20}.

\section{Acknowledgement}

We thank V. A. Khodel for fruitful discussions. This work was
partly supported by U.S. DOE, Division of Chemical Sciences, Office
of Basic Energy Sciences, Office of Energy. This work is partly
supported by the RFBR No. 19-02-00237.

\end{document}